\documentclass[twocolumn,prl,superscriptaddress,floatfix,aps,showpacs]{revtex4}
\usepackage{bm}
\usepackage{epsfig}
\usepackage{graphicx}
\usepackage{amsfonts,amssymb,amsmath}
\newcommand{\be}{\begin{equation}}
\newcommand{\ee}{\end{equation}}
\newcommand{\bfr}{\mathbf r}
\newcommand{\Delr}{ \Delta \bfr}
\newcommand{\bfk}{\mathbf k}

\begin{document}

\title{Off-Diagonal Long-Range Order in Solid $^{4}$He }
\author{Bryan K. Clark}
\affiliation{Dept. of Physics, University of Illinois at
Urbana-Champaign, Urbana, IL 61801, USA}
\author{D. M. Ceperley }
\affiliation{NCSA and Dept. of Physics, University of Illinois at
Urbana-Champaign, Urbana, IL 61801, USA}

\begin{abstract}
Measurements of the moment of inertia by Kim and Chan have found
that solid $^{4}$He acts like a supersolid at low temperatures. To
understand the order in solid $^4$He, we have used Path Integral
Monte Carlo to calculate the off-diagonal long range order (ODLRO)
(equivalent to Bose-Einstein condensation, BEC). We do not find
ODLRO in a perfect hcp crystal at the melting density. We discuss
these results in relation to proposed quantum solid trial
functions. We conclude that the solid $^4$He wave function has
correlations which suppress both vacancies and BEC.
\end{abstract}
%\maketitle

 \maketitle

Whether bulk solid has a region of its equilibrium phase diagram
with supersolid behavior has been of theoretical
interest\cite{andreev,chester,leggett} for many years. Recently,
Kim and Chan\cite{chan2} have observed behavior similar to that of
non-classical rotational inertia (NCRI) at temperatures below
0.2K. Analogous to behavior in a superfluid, NCRI would be
expected in a supersolid, a state of matter in which crystallinity
and superfluid behavior coexist. Two key theoretical quantities in
establishing the order of a bosonic system are the superfluid
fraction (estimated\cite{C181} to vanish) and off-diagonal long
range order (ODLRO) previously estimated only by variational
methods.

Off-diagonal order (equivalent to a non-zero condensate fraction)
tells us whether atoms at one end of the solid are in phase with
atoms at the other end of the solid and would supply a mechanism
for NCRI: for example, superfluid $^4$He in three dimensions has
both ODLRO and NCRI.
ODLRO\cite{penrose}, is a property of the single particle density
matrix:
\begin{eqnarray}
   n\left(\bfr,\bfr';\beta\right) &=& \frac{V}{Z}\sum_{\alpha}\int
d\bfr_{2}...d\bfr_{n} e^{-\beta
E_{\alpha}}\times\\
 \nonumber && \Psi_{\alpha}\left(\bfr,\bfr_{2}, \ldots ,\bfr_{n}\right)
 \Psi_{\alpha}^{*}\left(\bfr',\bfr_{2}, \ldots, \bfr_{n}\right)
 \end{eqnarray}
where $\Psi_{\alpha}$ and $E_{\alpha}$ are the $\alpha^{th}$
eigenfunction/value of the many body Hamiltonian, $Z$ the
partition function, $V$ the volume and $\beta=1/k_bT$. A system is
said to be bose-condensed if an eigenvalue of the matrix
$n(\bfr,\bfr')$ is proportional to the number of atoms in the
system. In this paper, we consider a translationally invariant
system in periodic boundary conditions. This implies that
$n(\bfr,\bfr')$ only depends on $\Delr=\bfr-\bfr'$ and its
eigenfunctions are momentum states. Considering the occupation of
the $\bfk=0$ momentum state, the condensate fraction is
 \be n_0= \frac{1}{V^2} \int d\bfr d\bfr' n(\bfr, \bfr'). \ee
A system has ODLRO if the condensate fraction is greater than zero
in the thermodynamic limit. That is:
$n_0=\lim_{\left|\Delr\right|\rightarrow\infty}n\left(\Delr\right)>0$.

Reatto\cite{reatto} showed that a pair-product (Jastrow) trial
wave function (JWF), commonly used for liquid $^{4}$He, has ODLRO.
This trial function has the form
$\Psi_J\left(R\right)=\exp(-\sum_{i<j} u\left(r_{ij}\right))$
where $u\left(r_{ij}\right)$ is fixed by minimizing the
variational estimate of the ground state energy. The Jastrow trial
function is equivalent to classical system with a pair potential
$v(r)$: the Jastrow factor maps into the classical pair potential
($v(r)\equiv 2 k_B T u(r)$). The physical argument for BEC is
straightforward: to calculate $n(\Delr)$ one displaces an
arbitrary atom by $\Delr$; if the average amplitude for this is
non-zero, the system has BEC. The condensate fraction will be
non-zero as long as $u(r)$ is short-ranged, except for special
systems such as hard spheres at close packing density. It will
also be true if the wave function has short-ranged three or higher
body correlation factors. Chester\cite{chester} then pointed out
that such a trial function can be a crystal, since it has the same
distribution as a classical solid interacting with a pair
potential. This argument can be used to prove\cite{proof} that
there exist quantum systems with both crystalline order and BEC.
Chester further conjectures that under the same conditions, a
crystal will have BEC only if there exist ground state vacancies
(or interstitials). This argument has been recently
elaborated.\cite{prokofev}

The Jastrow form of trial function turns out to give a seriously
incorrect density for freezing\cite{hansen} and other properties.
As is well-known, melting of classical systems  occurs when the
r.m.s. vibration around a lattice site equals about 0.14 of the
nearest neighbor distance: Lindemann's melting criterion. However,
solid $^4$He does not melt until much later\cite{burns}, with a
Lindemann's ratio of $0.30$. Assuming Lindemann's criteria for the
melting of a classical system, no Jastrow trial function can
describe a solid $^4$He near melting, nor can it describe a
quantum crystal without an intrinsic population of point defects.

A much better description of solid $^4$He is obtained by
multiplying the Jastrow function by a localization term; the
Jastrow-Nosanow form\cite{nosanow} (JN) or ``insulator'' form:
 \be
 \Psi_{JN} (R)= \frac{\Psi_J}{N!}\sum_P \prod_{i=1}^N \phi_{P_i}(\bfr_i).
 \ee
Here $\phi_i(\bfr)$ (a Wannier function) is localized about
lattice site $i$, and $P$ a permutation of atoms to lattice sites.
The sum over $P$ projects out the bosonic component of the wave
function; the projected wavefunction is a ``permanent''. In fact,
the symmetrization lowers the variational energy by an amount
proportional to the exchange energy, calculated in ref.
\cite{pimc} to be about 3 $\mu$K/atom; negligible for most
properties but crucial for properties such as BEC or NCRI. The JN
wave function is a very good description of the ground state
judging by the computed energies, Debye-Waller factors, pair
correlation function and estimated melting density\cite{hansen}.
However, the solid order is put in ``by hand'' rather than coming
about spontaneously. Broken translational symmetry is the hallmark
of the crystal state\cite{anderson}\cite{TI}. Upon freezing a
density wave arises which defines the lattice sites, a mean field
potential and hence $\phi_i(\bfr)$. It has been
shown\cite{schwartz} that this type of trial function does not
have BEC, assuming that the sum of the overlap of $\phi$ with
those on other sites is less than unity as is expected to be the
case. Because of the localized functions, the Reatto-Chester
theorem on BEC does not apply.

An alternative trial function\cite{lowy,C004, zhai} (we will
denote as a  metal, $\Psi_M$) is obtained by making the
single-body function $\phi(\bfr)$ independent of the lattice site,
but still having lattice symmetry. For electrons, $\phi(\bfr)$
would be a Bloch function of band theory and would be obtained as
a solution of a mean field with lattice symmetry. It is
expected\cite{andreev} that vacancy-interstitial (VI) fluctuations
play an important role in supersolidity. In the JN function, VI
pairs are bound, just as electron-hole pairs are in an electronic
insulator. In the metallic function, vacancies will be locally
attracted to interstitials but they are not bound as pairs; as a
result the $\Psi_M$ has both BEC and NCRI. Calculations on a
similar quantum solid\cite{C004} have found that $\Psi_M$ has a
higher energy than $\Psi_{JN}$; it costs energy to create unbound
VI fluctuations. We note that calculations\cite{pederiva} and
experiments\cite{simmons} on solid $^4$He suggest that there are
no unbound vacancies or interstitials at low temperatures.

Another trial function, the ``shadow'' wave
function\cite{vitiello} (SWF) has been introduced to allow a
translationally invariant trial function to have the correct solid
order. In this case, a single atom coordinate in the Jastrow trial
function is replaced by a composite object, a linear polymer,
with the inter- and intra- correlation factors being variationally
determined. Typically the polymer is in fact a dimer, with the two
ends separated by a distance on the order of the inter-atomic
spacing. One of the ends (the shadow) is integrated over to get
the trial function. The great advantage of the SWF is that the
crystal order is spontaneously generated, and one gets accurate
properties after optimizing the wave function parameters.  As more
monomers are added to the trial function, the functional form is
equivalent to the Feynman-Kacs path integral expression for the
ground state wave function\cite{spigs}.   Though it has not been
analytically shown, simple arguments involving displacing the
polymer, make it is plausible that the shadow trial function will
have always have BEC, as long as the inter-atomic correlation
factors are short-ranged. This has been verified in recent
numerical calculations by Galli et al.\cite{galli}

Thus, we have a dilemma, not uncommon with arguments based on
variational wavefunctions: one can have several satisfactory trial
functions, all of them capable of good descriptions of solid
helium, but some are BEC and some are not BEC. All of the usual
properties one uses to test the quality of the trial function,
{\it e. g.} the energy, the Debye-Waller factor, the pair
correlation function, etc. are diagonal in coordinate space
(expectations of $|\Psi(R)|^2$); hence they are unreliable
measures of how accurate the off-diagonal elements needed in Eq.
(1) are. As we have mentioned above, it is not possible to state a
general theorem covering whether all quantum crystals must or
cannot have BEC, since the Jastrow trial functions can be used to
define a Hamiltonian which is supersolid.
We need a more direct, reliable method to decide whether the
ground state of solid $^4$He is BEC and, hence, which class of
trial function, metal or insulator, is appropriate.  Calculating
$n(\Delr)$ in solid bulk $\,^{4}$He allows us to establish how
good these trial functions are for computing BEC in quantum
crystals.

To provide an unbiased answer, we use path integral Monte Carlo
(PIMC)\cite{pimc}, a numerical method that calculates integrals
over the many body density matrix. It is ideally suited for this
calculation since it can be done at finite temperature (under
conditions where an experimental signature of NCRI has been seen),
is, in principle, exact, and has been validated on many
properties\cite{pimc} of liquid and solid $^4$He. Most
importantly, it is independent of a trial wave function bias or
any assumption of lattice.  Only the He-He potential enters: a
semi-empirical form\cite{aziz} is known to be accurate; in any
case, experimental results suggest that supersolid behavior is a
robust phenomenon insensitive to fine details of the interaction.

In PIMC, based on Feynman's description of superfluid helium, one
maps the quantum system onto a classical system of ``polymers''
which can permute onto one another. The value of
$n\left(\left|\Delr\right|\right)$ is mapped to the end-to-end
distribution of an open ``polymer'' in a sea of closed
``polymers'' representing the other helium atoms. If the system
has ODLRO, the two ends will become separated from each other; the
condensate fraction is the value of $n(\Delr)$ at large $r$. On
the other hand, if there is no ODLRO, the two ends will remain
localized relative to each other and $n(\Delr)$ will approach $0$
at large $|\Delr|$. Superfluidity ({\it i.e.} NCRI), on the other
hand, is represented in PIMC as long cyclic permutation of paths,
for example, as a complete row of atoms shifting down one element.
One can immediately see the difference between ODLRO and NCRI in a
PIMC solid calculation. Suppose we have N atoms and N lattice
sites, so each site is, on the average, singly occupied. To
perform the ODLRO calculation, one of the atoms paths is opened
up, so now there will be $N-1$ closed polymers and 2 open ends;
clearly one of the $N$ sites must be doubly occupied if we count
the open ends as distinct atoms. For NCRI one can simultaneously
shift all atoms to their new permuted sites; there need not be any
double occupation, but such a simultaneous shift is unlikely for
many atoms.

Techniques\cite{pimc} developed and tested\cite{timestep} on
superfluid $^4$He  are used to calculate $n(\Delr)$ efficiently,
although new moves are used to ensure ergodicity in the solid. If
one simply cuts open one polymer, one rarely finds large values of
$\Delr$. To guarantee sufficient statistics on the end-end
distance, we multiply the path integral density by a series of
importance sampling functions having the effect of
forcing apart the ends. A series of importance functions are used
to get good statistics for a specific range of magnitudes of
$|\Delr|$ The separate end-to-end distributions are then
normalized with respect to each other by minimizing the $\chi^2$
difference between them and using the definition $n(0)=1$.
We perform calculations of $n(\Delr)$ both by performing an
angular average over $\Delr$ as well as fixing $\Delr$ in the
basal plane, specifically in the $x$ direction. Calculations of
$n(\Delr)$ can much more easily access the full path space than
calculations of the superfluid density because the ends of the
open polymer can move more freely.

Calculations have been performed in a nearly cubic box of 180
particles as well as a series of longer rectangular boxes extended
in the basal plane ranging from 48 to 144 atoms. For the longer
box, we altered our importance function to force the open ends
apart only in the $x$ direction (in the nearest neighbor direction
of the basal plane).
We examine values of $n(\Delr)$ for a range of temperatures from
$0.1K$ to $2K$ in a (periodic) rectangular box of size $\sim
18.5~\AA$ corresponding to a density, 0.0286$~\AA^{-3}$ near the
experimental melting pressure of 25.3 bars.  The random walk is
started with the atoms in a perfect hcp crystalline lattice but no
constraints are placed on how the atoms can arrange themselves.
Since the box is chosen to be commensurate with a perfect hcp
crystal, vacancies and interstitials are expected not to exist in
the system though they can form by a fluctuation. We have verified
that the results are independent of temperatures below $2.0K$.

The results of these calculations for $T=0.5K$ are shown in figure
1. Note the curvature at small $\Delr$ is determined by the kinetic
energy estimated at 24.0 K/atom.
  For both the spherical average value and the value in
the basal plane $n\left(\left|\Delr\right|\right)$ decays
exponentially for $|\Delr| >3 \AA$: we find no indication of ODLRO
(BEC) in bulk solid $^{4}$He.
In fact the rate of decrease for $n(x)$, the SPDM in the x-direction,
is in agreement with the frequencies
for straight line winding exchange\cite{C181}, shown as the solid
line.
We find oscillations in the computed $n(x)$
reflecting the underlying crystal lattice.  The angular averaged
$n(r)$ shows less structure than that in the x-direction. Though
it is possible that $n(x)$ will plateau for a longer distance
scale, we see no indication of this in the results, nor can we
think of a physical mechanism that would be
responsible\cite{smallbec}.
A system with ODLRO will be manifested when one end of the open
polymer loses knowledge about the other end of the open polymer.
But we find that pulling apart the two ends takes ``work'' per
unit length, no matter how far apart the two ends are. This
happens because additional atoms are displaced from their lattice
sites.

The results of our PIMC calculations are in quantitative agreement
with variational Monte Carlo calculations using a shadow wave
function\cite{galli} for distances less than 9 $\AA$; they
conclude that BEC does exist in the ground state of $^{4}$He with
a very small condensate fraction, $n_0=5 \times 10^{-6}$ at the
melting density. However, we find that $n(\Delr)$ continues to
decreases beyond 9 $\AA$.

We believe the reason for the discrepancy with PIMC is that the
shadow wave function is a technique that improves upon the Jastrow
wave function, in fact it is remarkably accurate up to the second
neighbor distance. Nonetheless the wavefunction is still built
from short range correlations. This leads inevitably to a non-zero
superfluid fraction. We also expect the SWF will give a non-zero
but very small vacancy concentration\cite{vacancy}.
PIMC does not make a variational ansatz, with the sole assumption
that the results at low temperature smoothly approach the ground
state values.

Among the proposed quantum solid wave functions, the
symmetry-breaking Nosanow-Jastrow wavefunction, gives results in
agreement with PIMC, validating the ``insulator'' representation
of solid helium. (Further details about this picture are in ref.
\cite{anderson}). The implication of our calculation and the
Reatto-Chester theorem is that a quantum solid trial function must
include long-range correlations. This is most simply done by
putting in localized functions about lattice sites, reflecting the
broken symmetry. We have only done PIMC calculations at the
melting density but we do not expect different behavior at higher
density, since difficulty of exchange grows rapidly with density.
The result reported here, together with the finding of zero
superfluid density reported in ref. \cite{C181} suggest that the
mechanism for the measurements of Kim and Chan\cite{chan2}
involves more than equilibrium properties of a perfect $^4$He
crystal.

\begin{figure}
\begin{center}
\epsfig{file=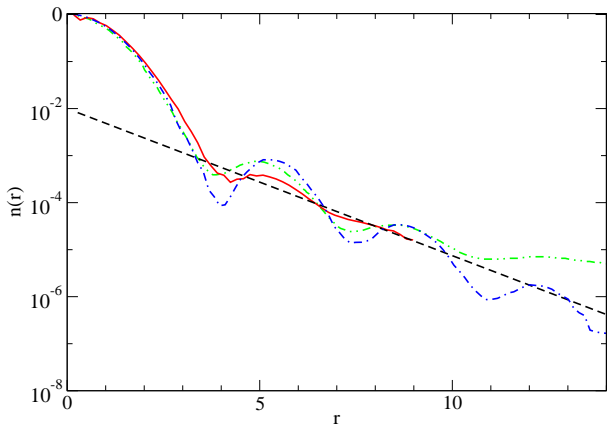,width=0.4\textwidth,angle=-90}
\caption{(color online) The single particle density matrix as a
function of distance (in $\AA$), estimated with PIMC, in hcp
$^4$He at a density 286$nm^{-3}$ and a temperature of 0.5K. The
red curve is the spherically averaged n(r) while the blue curve is
n(x). The upper (green) curve is a variational MC calculation
using the shadow wave function ref. \cite{galli}.  The straight
line (black) has a slope determined from the exchange
calculations of ref. \cite{C181}. } \label{spdm}
\end{center}
\end{figure}

(After the calculation reported here was completed, we have
learned of a similar PIMC calculation \cite{boninsegni}.  The
overall conclusions are similar, however their Monte Carlo
algorithm  and results for $n(\Delr)$ differ in important ways.)

This work was supported by NSF under grant no. DMR-03 25939 ITR
and the fundamental physics program at NASA (NAG-8-1760). Computer
time has been provided by NCSA and the Materials Computation
Center and the Frederick Seitz Materials Research Laboratory (U.S.
DOE DEFG02-91ER45439) at the University of Illinois
Urbana-Champaign. We acknowledge useful communications with
D.Galli, L. Reatto, W. Saslow, N. Trivedi and G. V. Chester.


\begin{thebibliography}{21}
\bibitem{proof}To prove that there exist supersolids, assume that the exact
bosonic wavefunction equals the potential of a known classical
crystal, e.g. a Lennard-Jones crystal\cite{stillinger}. Then
define the Hamiltonian of the quantum system to have this as an
exact ground state; it must be the ground state since it is
nodeless. The quantum potential as defined is not a pair
potential; it has a very strong 3-body potential. The crystal will
also have a non-zero density of vacancies since the equivalent
classical system does.
\bibitem{TI}Although it is true
that the exact wave function for a finite system will have
translation invariance (TI), a trial function lacking TI will be
closer to the thermodynamic limit since if one averages over the
center-of-mass, that lowers the energy by order (1/N), but such
motion is unphysical in the limit of large N.
\bibitem{timestep}We used pair density matrices and found an
imaginary-time step of $0.0125 K^{-1}$ was sufficient for
convergence.
\bibitem{smallbec}The observed transition temperatures  rule out very small condensate fractions since
one expects that if ODLRO is due to BEC of point defects then
$n_0\sim \hbar^3(mk_BT_c)^{3/2} \sim 0.01$.
\bibitem{vacancy}The concentration of vacancies is given directly by the
square of the trial wavefunction, $|\Psi|^2$ in the thermodynamic
limit, e.g.  considered as a classical Boltzmann
distribution\cite{stillinger}. For approximate trial functions,
this will give a different answer than a variational estimate of
the vacancy energy from the Hamiltonian: $E_{v}(N\pm
1,N)-E_{v}(N,N)$ as was done in ref. \cite{swf_vacancy}.  Here,
$E_{var} (N,N_s)$ is the variational energy of $N$ atoms and $N_s$
lattice sites at constant atomic density.)
\end{thebibliography}
\end{document}